\documentclass[aps,pre,twocolumn,groupedaddress,showpacs]{revtex4-1}

\usepackage{amsmath}
\usepackage{amssymb}
\usepackage{graphicx}
\usepackage{color}
\usepackage{amsfonts,amsbsy,eucal,dsfont}
\usepackage{array}

\newcommand{\erf}{\operatorname{erf}}

\begin{document}

\title{Flow of colloidal solids and fluids through constrictions: dynamical density functional theory versus simulation}
  \author{Urs Zimmermann}
  \email{Urs.Zimmermann@uni-duesseldorf.de}
  \author{Frank Smallenburg}
  \author{Hartmut L\"owen}
  \affiliation{Institut f\"ur Theoretische Physik II: Weiche Materie, Heinrich-Heine-Universit\"at D\"usseldorf, D-40225 D\"usseldorf, Germany}
\date{\today}

\begin{abstract}
Using both dynamical density functional theory and particle-resolved Brownian dynamics simulations, we explore the flow of two-dimensional colloidal solids and fluids 
driven through a linear channel with a geometric constriction. The flow is generated by a constant external force acting on all colloids. 
The initial configuration is equilibrated in the absence of flow and then the external force is switched on instantaneously. 
Upon starting the flow, we observe four different scenarios: a complete blockade,
a monotonic decay to a constant particle flux (typical for a fluid), a damped oscillatory behaviour in the particle flux,
and  a long-lived stop-and-go behaviour in the flow (typical for a solid).
The dynamical density functional theory describes all four situations but predicts infinitely long undamped oscillations in the flow
which are always damped in the simulations. We attribute the mechanisms of the underlying stop-and-go flow to
symmetry conditions on the flowing solid. Our predictions
are verifiable in real-space experiments on magnetic colloidal monolayers
which are driven through structured microchannels and can be exploited to steer
the flow throughput in microfluidics.
\end{abstract}

\maketitle

\section{Introduction}

Particle flow through constrictions occurs in widely different situations ranging from nanofluidics \cite{EijkelvdB2005,SchochHR2008,BocquetC2010}
to medicine \cite{FallahEtAl2010,NoguchiGSWF2010,BernabeuNGCHKC2013}
and crowd management \cite{PeacockKA2011}.
On the nanoscale, the permeation of molecules through pores is
controlled by constrictions \cite{DzubiellaH2005}.
On the mesoscale, colloidal suspensions \cite{WyssBMSW2006,GenoveseS2011,KreuterSNLE2013},
dusty plasmas \cite{IvlevLMR2012}, and micron-sized bacteria \cite{HulmeDLSTMBW2008,AltshulerMPPRLRC2013} passing through micro-patterned
channels as well as vascular clogging by parasitized red blood cells \cite{PatnaikDMMSM1994} are important examples.
Finally, in the macroscopic world, granulate fluxes  through silos \cite{ZuriguelJGLAM2011,ThomasD2015,Behringer_arXiv_2015}
and the escape of pedestrians or animals through narrow doors \cite{HelbingFV2000b,KirchnerNS2003,ZuriguelEtAl2014,GarcimartinPFRMGZ2015}
illustrate the relevance of constricted flow phenomena. 

Despite its relevance, flow through geometric constrictions
is still not understood from a non-equilibrium statistical physics point of view
within a fundamental microscopic theory.
Classical density functional theory (DFT) \cite{Evans1979,Loewen1994a, Singh1991, TarazonaCMR2008, Roth2010} constitutes
such a microscopic approach in equilibrium. In principle, 
DFT can be used to calculate the equilibrium phase diagram - including the freezing and melting lines - 
for given interparticle interactions and thermodynamic conditions (such as prescribed temperature and chemical potential). 
This is done by minimizing the appropriate free-energy functional with respect to the one-particle density distribution, which captures the structural properties of each phase.
Although the theory is in practice approximative, as the exact functional is not known, there are very good
approximation schemes (e.g. for hard spheres and hard disks) with remarkable predictive power
\cite{Roth2010,Oettel_PRE2010, HaertelOREHL2012}. A geometric constriction can be conveniently modelled by
an external curved wall, a set-up which can directly be accessed by density functional theory.
Particle flow, however, is a non-equilibrium situation, such that standard equilibrium DFT cannot be applied
directly.
For completely overdamped Brownian dynamics, i.e. for mesoscopic colloidal particles in a solvent, it was
shown that DFT can be generalized to describe the non-equilibrium relaxation dynamics of the
time-dependent one-particle density \cite{MarconiT1999,ArcherE2004,EspanolL2009,WittkowskiLB2010}.
The resulting dynamical density functional theory (DDFT) has been
applied to a variety of non-equilibrium phenomena. 
These include  colloids in external  shear fields
such they are advected by the solvent flow \cite{RauscherDKP2007,Brader_Krueger_MolPhys_2011,Krueger_Brader_EPL_2011,Reinhardt_EPL_2013}, microrheology 
where a particle is driven through a colloidal background
\cite{AlmenarR2011}, solvent-mediated hydrodynamic interactions
\cite{RexL2008,RexL2009,Rauscher2010,Goddard_MMS_2012,Goddard_PRL_2012,GoddardNSYK2013,Goddard_JCP_2013,Donev_JCP_2014},
diffusion in hard sphere fluids at high volume fractions \cite{Stopper_PRE_2015}
and in binary mixtures \cite{Malijevsky_JCP_2013}, feedback control of colloids
\cite{Lichtner_PRE_2012} and the  collapse of a
colloidal monolayer as governed by attractive interactions \cite{Bleibel_SM_2014}.
Moreover colloidal crystal growth \cite{vanTeeffelenLL2008,vanTeeffelenBVL2009,NeuhausNJP2013,NeuhausPRE2013} and quasicrystal growth
\cite{Achim_PRL_2014,Archer_PRE_2015} (see Ref. \cite{Nagao_PRL_2015} for a recent experiment) have been tackled by DDFT-like approaches.
  Finally, active colloids \cite{WensinkL2008,WittkowskiL2011,MenzelL2013,Menzel_Saha_2015}
and even granulate dynamics
\cite{Marconi_JCP_2007,MarconiTCM2008,MarconiM2009,MarconiM2010,Marconi_CommTP_2014}
have been described using DDFT.

In this paper, we apply DDFT to the flow of Brownian particles through a geometric constriction.
This is realized by colloidal particles flowing through microchannels \cite{KreuterSNLE2013,GenoveseS2011}.
Here we restrict ourselves to two spatial dimensions and consider the flow of colloids through a
structured channel as motivated by experiments of superparamagnetic colloids in two dimensions
\cite{Straube_Dullens_EPL_2011,Juniper_NatCom_2015}.
We use an equilibrium density functional for two-dimensional parallel dipoles similar to earlier work \cite{vanTeeffelenLHL2006},
which reproduces the fluid-solid transition in two dimensions. We then employ DDFT to describe a flow situation
in a linear channel where particles are driven by a constant external force. The channel includes a constriction,
where the channel gets narrower. We systematically explore the influence of this constriction on the net particle flow, using both DDFT and
Brownian dynamics computer simulations. In both methods, we equilibrate the system in the absence of flow, and measure the time-dependent flow through
the constriction after instantaneously switching on the external driving force.

Within DDFT we find that the averaged flow through the constriction is qualitatively different for solids and fluids:
in the fluid the flow is constant (i.e. time-independent) while in the solid it is
periodically oscillating as a function of time. This interesting intermittent flow
is induced by the constriction as it vanishes in the pure linear channel in the absence of any constriction.
Therefore it is not a trivial passing of particle layers but rather a self-organized oscillation
generated by the constraint breaking the one-dimensional translation symmetry along the channel.
The computer simulations corroborate the theoretical findings qualitatively insofar as a different
behavior is revealed in the time-dependent flow in the solid and in the fluid.
For solids there is an intermittent flow with damped oscillatory correlations in time while for fluids
these oscillations are overdamped. This can be expected as DDFT is a mean-field theory
which averages in a global and approximative sense, while the simulations contain explicit stochastic noise,
responsible for damping the oscillatory behavior.

In more detail, depending on the initial state (fluid or solid) and on the width of the geometric constriction, we identify
four different situations: i) a complete blockade on the time scale of the calculations,
ii) a monotonic convergence to a constant particle flux (typical for a fluid), iii) strongly damped oscillations in the particle flux,
and iv)  a long-lived stop-and-go behaviour in the flow (typical for a solid).
We attribute the underlying stop-and-go flow to symmetry conditions on the flowing solid by studying the 
case of five and six crystalline layers as an example.
Our predictions are verifiable in real-space experiments on magnetic colloidal monolayers
which are driven through structured microchannels, e.g. by gravity. They can further be exploited to steer
the flow throughput in microfluidics and to tailor the pouring of colloidal particles through nozzles.

The paper is organized as follows: In section \ref{sec:model} we describe the details of the system under investigation. In section \ref{sec:ddft}
the dynamical density functional theory approach is presented and in section \ref{sec:bd} we describe the computer simulations.
Results of both methods are presented and discussed in section \ref{sec:results}. Our conclusions are presented in section \ref{sec:conclusion}.

\section{The Model}\label{sec:model}
\subsection{Interaction}
We consider point--like Brownian particles in two spatial dimensions which interact via a pairwise potential 
\begin{equation}
        u(r) = \frac{u_0}{r^{3}},\label{eq:interaction1}
\end{equation}
where $r$ is the distance between two particles and the amplitude $u_0 > 0$ sets the interaction strength.
A real--world analogue of this system is given by superparamagnetic particles
that are confined in a 2--d plane with an uniform external magnetic field
$\mathbf{B}_\text{ext}$ applied perpendicular to the plane.
The external magnetic field $\mathbf{B}_\text{ext}$ induces a dipole-dipole interaction between the colloidal particles, which can be tuned
by changing its strength. In bulk, the only relevant length scale present in this 
system is the typical interparticle distance, which is given by 
\begin{equation}
l = \rho_0^{-1/2},
\end{equation}
with
\begin{equation}
\rho_0 = N/A_0
\end{equation}
the number density of the system, $N$ the number of particles, and $A_0$ the accessible area, which will be defined later. Due to the 
inverse power law scaling of Eq. \ref{eq:interaction1}, a change in density of the system is equivalent to a change in the 
interaction strength $u_0$. It is therefore convenient to rewrite Eq. \ref{eq:interaction1} as
\begin{equation}
\frac{u(r)}{k_\mathrm{B} T} = \frac{\Gamma}{(r/l)^3} \label{eq:interaction}
\end{equation}
where $\Gamma = u_0\rho_0^{3/2} / (k_\mathrm{B} T)$ is a dimensionless coupling parameter. 
The bulk phase behavior of these particles is characterized by a fluid at low $\Gamma \lesssim 11$, and a hexagonally ordered solid phase at high $\Gamma \gtrsim 12$ \cite{kalia1981interfacial,haghgooie2004structural}.

Naturally, this phase diagram is expected to change significantly in the confinement of a channel, as considered here. In particular, as the system is effectively one-dimensional, we expect only short-range ordering in the channel, and no true fluid to crystal transition. Nonetheless, at high $\Gamma$ we do expect local ordering into a hexagonal lattice, aligned with the boundaries of the channel \cite{haghgooie2004structural}.

\subsection{Channel confinement}
\begin{figure}
 \begin{tabular}{m{0.29\textwidth}m{0.19\textwidth}}
a)&b)\\
\includegraphics[width=0.28\textwidth]{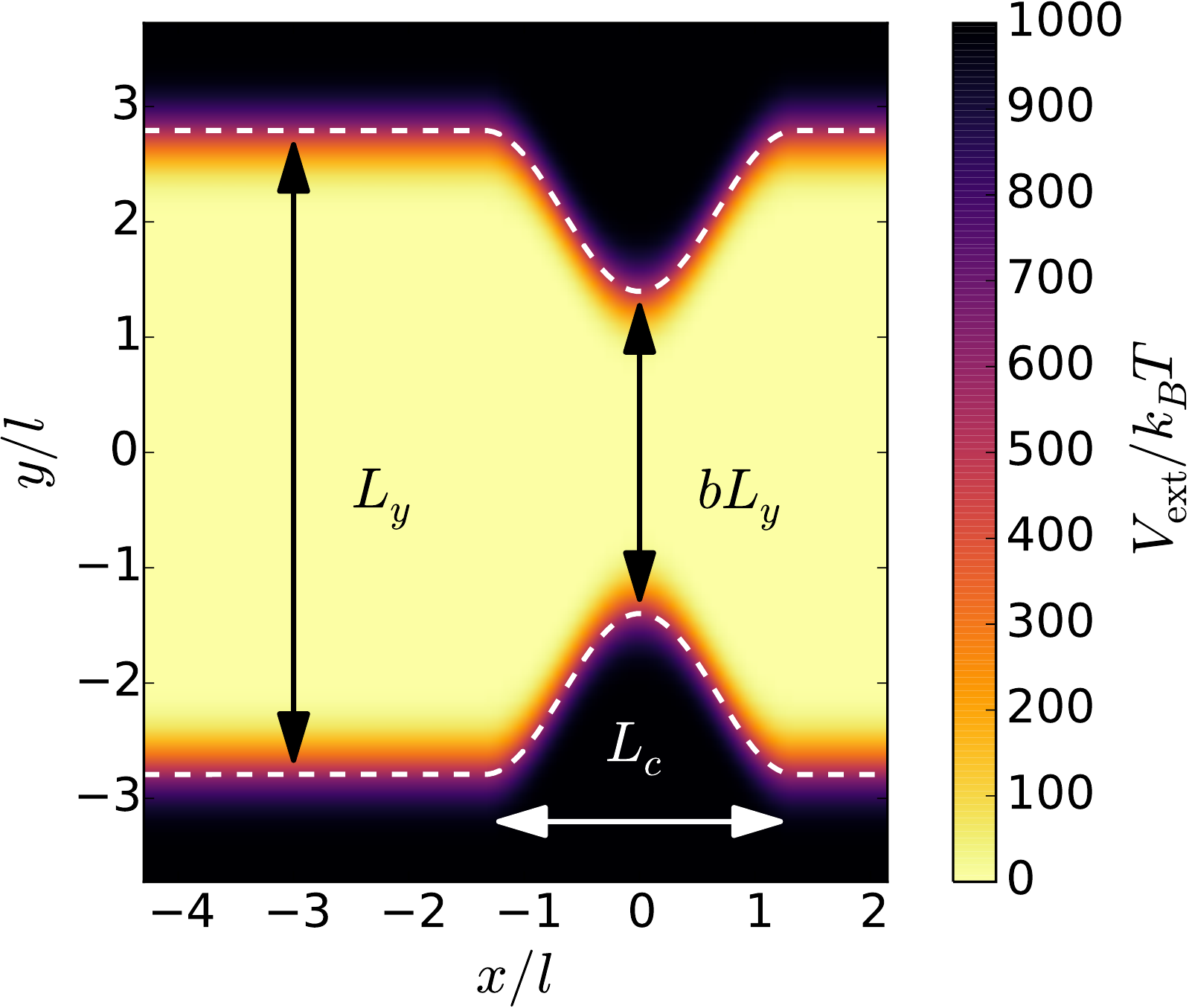}&\vspace{-1cm}\includegraphics[width=0.18\textwidth]{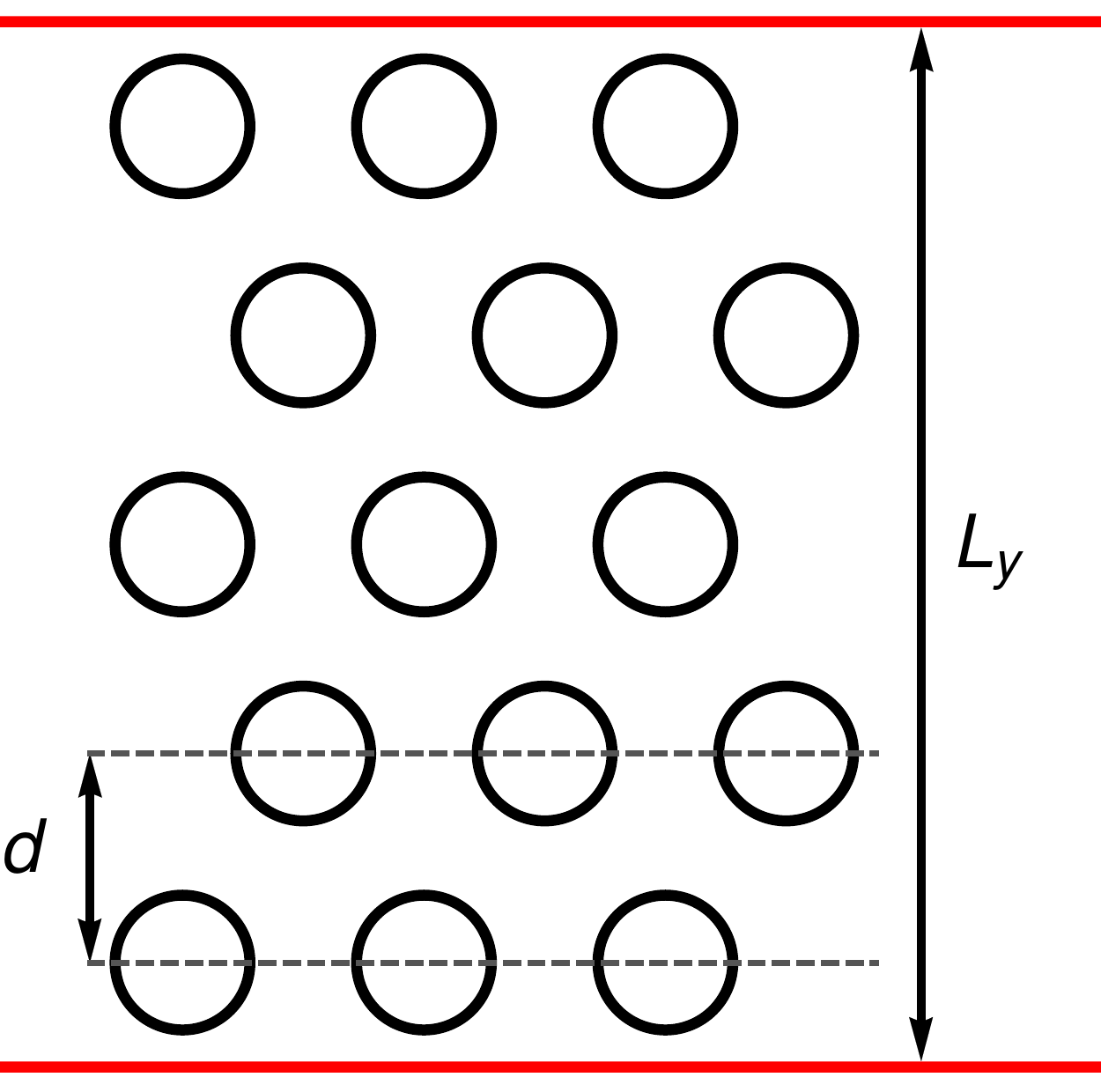}
\end{tabular}
\caption{{\bf (a)} Potential energy $V_\mathrm{ext}(x,y)$ in the channel, for $b=0.5$, $L_c=2.686l$, $L_y = 6d$, $w=0.25l$ and $x_0=0$. The dashed lines represent $\pm g(x)$ and enclose the accessible area $A_0$. {\bf(b)} Schematic representation of the channel dimensions and the typical hexagonal lattice observed within the channel at high $\Gamma$. Note that $d$ is defined in a perfect hexagonal lattice and may vary in the channel.}
\label{fig:system}
\end{figure}

Inside the 2--d plane the particles are additionally confined in a channel geometry along the $x$--axis, represented by an
external potential $V_\text{ext}(x,y)$. The lateral profile of the channel is modeled as error--function steps at the walls of the channel so the external potential is given by
\begin{eqnarray}
    V_\text{ext}(x, y) = V_0 \left[1 - \frac{1}{2} 
                 \erf\left(\frac{y + g(x)}{\sqrt{2}w}\right)\right. \nonumber\\
  +  \frac{1}{2}\left. \erf\left(\frac{y - g(x)}{\sqrt{2}w}\right) \right],
\end{eqnarray}
with $V_0$ being the maximum potential height, $\pm g(x)$ describing the contour lines of the channel walls and $w$ characterizing the softness of the walls.
For a straight channel with width $L_y$ and without constriction the contour functions are simply
$g(x) \equiv \frac{L_y}{2}$.
The constriction is modelled as a single cosine wave of length $L_c$ at $x_0$ that is added smoothly to
the channel contour. Therefore, $g(x)$ is given by
\begin{align}
    g(x) = \begin{cases} \frac{L_y}{2} - \alpha\left[1 + \cos\left(2\pi\frac{x-x_0}{L_c}\right)\right], \text{ if } \left|x - x_0\right| < \frac{L_c}{2}\\
                 \frac{L_y}{2}, \text{ otherwise}  
                     \end{cases}
\end{align}
with amplitude $\alpha=\frac{L_y}{4}(1 - b)$.
Here, we introduced the parameter $b$ as the ratio of constriction width over the total channel width. Consequently, $0 \leq b \leq 1$, where $b=0$ refers
to a completely blocked channel and $b=1$ is a channel without constriction.
See Fig. \ref{fig:system}a for an illustrative sketch of $V_\text{ext}(x, y)$.

We define the accessible area as the region between the midlines of the two walls, i.e.
\begin{equation}\label{eq:area}
        A_0 = 2\int_{-L_x/2}^{L_x/2} g(x) dx.
\end{equation}

By definition, the number density in the system is given by $\rho_0 = N/A_0 = 1 / l^2$, with $l$ our unit of length.
In this work, we focus on channels with a width chosen such that either five or six crystalline layers reliably form within the channel, 
oriented such that lines of nearest-neighbors are aligned with the channel walls (see Fig. \ref{fig:system}b). However, the number of defects 
in this crystal strongly depends on the commensurability between the channel width and the lattice spacing of the crystal \cite{haghgooie2004structural}. 
In a perfect hexagonal lattice at density $\rho_0 = 1/l^2$, the distance between two crystal layers is
\begin{equation}
 d =  \sqrt{\frac{\sqrt{3}}{2}}l,
\end{equation}
and we will adopt this definition
of $d$ for our confined system as well.
In order to accomodate a crystal with a low number of defects, we therefore choose the channel width to be $L_y = n d$, with $n= 5$ or $6$. 
Both DDFT and simulations show that this indeed leads to crystals with the desired number of layers.

In order to further reduce parameter space, we fix the constriction length $L_c = 2.686 l$, wall softness $w=0.25 l$ and $V_0 = 1000 k_B T$. 

\subsection{Equations of motion}
We model the dynamics of the particles in the channel via simple, overdamped Brownian dynamics. The equations of motions are given by:
\begin{equation}
        \dot{\mathbf{r}}_i = \frac{D}{k_B T} \mathbf{F}_i(\mathbf{r}^N) + \sqrt{2D}{\boldsymbol \xi}_i(t),\label{eq:eom}
\end{equation}
where $\mathbf{r}_i$ are the coordinates of the $i$th particle and $\mathbf{r}^N \equiv (\mathbf{r}_1, \dots, \mathbf{r}_N)$ is a short-hand notation for the coordinates of all particles, $D$ is the diffusion constant of a single particle without external forces, $\mathbf{F}_i(\mathbf{r}^N)$ is the total force
acting on the $i$th particle composed of pair interactions, external potential, and dragging force:
\begin{equation}
 \mathbf{F}_i = -\sum_{j\neq i} \nabla_i u(\left|\mathbf{r}_i - \mathbf{r}_j\right|) - \nabla_i V_\text{ext}(\mathbf{r}_i) + f \hat{\mathbf{x}},
\end{equation}
with $u(r)$ given by Eq. \ref{eq:interaction} and $\nabla_i$ being the gradient operator with respect to particle coordinates $\mathbf{r}_i$ and the unit vector in $x$-direction $\hat{\mathbf{x}}$. The external force responsible for the flow of particles through the channel is modeled via a constant force $f$ along the $x$--axis.
Finally, ${\boldsymbol \xi}_i(t)$ is a delta-correlated Gaussian noise process modeling the thermal fluctuations. In the remainder of this work, we will fix the drag force $f = 1 k_B T / l$. As a unit of time, we will use the time it takes a particle to diffuse by a typical distance of $l$, i.e.
\begin{equation}
 \tau = \frac{l^2}{D}.
\end{equation}

A stochastically equivalent description of Eq. \ref{eq:eom} is given by the Smoluchowski picture in which the time--dependent $N$--particle probability distribution $p(\mathbf{r}^N, t)$ is considered. The Smoluchowski equation is given by
\begin{equation}\label{eq:Smoluchowski}
 \frac{\partial p(\mathbf{r}^N,t)}{\partial t} = D \sum_{i=1}^{N} \nabla_i [k_\text{B}T \nabla_i + \mathbf{F}_i] p(\mathbf{r}^N,t).
\end{equation}
An integration over the probability distribution $p(\mathbf{r}, t)$ with respect to all but one coordinate gives the one--particle density
\begin{equation}\label{eq:one_particle_density}
 \rho(\mathbf{r}_1, t) = N \int \! \text{d}\mathbf{r}_2 \dots \int \! \text{d}\mathbf{r}_N p(\mathbf{r}^N, t),
\end{equation}
which describes the ensemble averaged particle density at time $t$ and is the basic quantity in the DDFT.

\section{Dynamical Density Functional Theory}\label{sec:ddft}
\subsection{General Theory}
Dynamical Density Functional Theory 
(DDFT) is conveniently derived from the Smoluchowski equation (\ref{eq:Smoluchowski}) by projecting onto the one-particle density
and invoking the additional adiabatic approximation \cite{ArcherE2004}. As a result, DDFT is an approximative theory. It can be written 
as a continuity equation
\begin{align}\label{DDFT}
    \frac{\partial \rho (\mathbf{r},t)}{\partial t} &= D \nabla \left(\rho(\mathbf{r},t) \nabla \frac{\delta \mathcal{F}[\rho(\mathbf{r},t)]}{\delta \rho(\mathbf{r},t)}\right),
\end{align}
which expresses the particle number concentration $\rho (\mathbf{r},t)$. The current density $\mathbf{j}(\mathbf{r},t)$ is explicitly given by a generalized Fick's law:
\begin{align}\label{eq:ddft_flux1}
 \mathbf{j}(\mathbf{r}, t) = - D \rho(\mathbf{r}, t) \nabla \frac{\delta \mathcal{F}[\rho(\mathbf{r},t)]}{\delta \rho(\mathbf{r},t)},
\end{align}
with the Helmholtz free energy functional 
\begin{equation}
\mathcal{F}[\rho] = \mathcal{F}_\text{id}[\rho] + \mathcal{F}_\text{ext}[\rho] + \mathcal{F}_\text{exc}[\rho]
\end{equation}
which can be split in three principal contributions.
The ideal gas term 
\begin{equation}
\mathcal{F}_\text{id}[\rho] = k_\text{B}T \int \! \text{d}\mathbf{r} \, \rho(\mathbf{r},t) \big( \log(\Lambda^2 \rho(\mathbf{r},t)) - 1 \big)
\end{equation}
and the external potential contribution 
\begin{equation}
\mathcal{F}_\text{ext}[\rho] = \int \! \text{d}\mathbf{r} \, \rho(\mathbf{r}) (V_\text{ext}(\mathbf{r}) + fx)
\end{equation}
with thermal de Broglie wavelength $\Lambda$ are known expressions.
In contrast, the excess free energy functional $\mathcal{F}_\text{exc}[\rho]$, which describes the particle interactions,
is unknown and has to be approximated. Here, we use the Ramakrishnan--Yussouff functional described in the next subsection. Substituting the first two terms in Eq. \ref{eq:ddft_flux1}, the current is thus given explicitly by
\begin{align}\label{eq:ddft_flux}
 \mathbf{j}(\mathbf{r}, t) = -D\nabla\rho(\mathbf{r}, t) + \rho(\mathbf{r}, t) \nabla \big(V_\text{ext}(\mathbf{r}) - fx + \frac{\delta \mathcal{F}_\text{exc}[\rho]}{\delta \rho(\mathbf{r}, t)} \big).
\end{align}
Since we are only interested in the flux along the channel, we define the particle flow in the $x$-direction,
i.e. 
\begin{equation}
j_x(x,t) = \int_{-\infty}^\infty \! \mathrm{d}y \; \mathbf{j}(\mathbf{r},t) \cdot \hat{\mathbf{x}}
\end{equation}
The average flow through the channel $\bar{j}_x$ can then simply be defined as the long-time average value of $j_x(x,t)$:
\begin{equation}
\bar{j}_x = \lim_{T\to\infty} \frac{1}{T}\int_0^T \! \mathrm{d}t \; j_x(x,t).
\end{equation}
Note that as the particle density is a conserved quantity, $\bar{j}_x$ is independent of the position $x$.

\subsection{Excess Functional}
We chose the Ramakrishnan--Yussouff expression \cite{RamakrishnanY1979} as an approximate excess free energy functional, which is a convenient
way to model soft and long-ranged particle interactions. The functional derivative of the Ramakrishnan--Yussouff functional
is given as a convolution of $\rho(\mathbf{r},t)$ and the pair (two-point) direct correlation function $c^{(2)}_0(r;\rho_0, \Gamma)$ of an isotropic and
homogeneous reference fluid with the prescribed density $\rho_0 = 1/l^2$, at interaction strength $\Gamma$:
\begin{align}\label{eq:RY}
 \frac{\delta \mathcal{F}_\text{exc}[\rho(\mathbf{r},t)]}{\delta \rho(\mathbf{r},t)} = -
 k_\text{B}T \int \!\! \text{d}\mathbf{r'} \;\rho(\mathbf{r'},t) c^{(2)}_0(|\mathbf{r} - \mathbf{r'}|; \rho_0, \Gamma).
\end{align}
We use the direct correlation functions obtained by liquid integral theory with the Rogers-Young closure which were calculated in Ref. \cite{vanTeeffelenLL_JPCM_2008}, where it was shown that despite its simplicity the Ramakrishnan--Yussouff functional accounts for the freezing transition in two dimensions at $\Gamma \gtrsim 36.2$. 

Since the functional derivative of the excess functional Eq. \ref{eq:RY} is a convolution of $\rho(\mathbf{r},t)$ and $c^{(2)}_0(r;\rho_0, \Gamma)$ we can efficiently compute its value using fast Fourier transform.

\subsection{Protocol}

\begin{figure}
\vspace{0.5cm}
\includegraphics[width=0.45\textwidth]{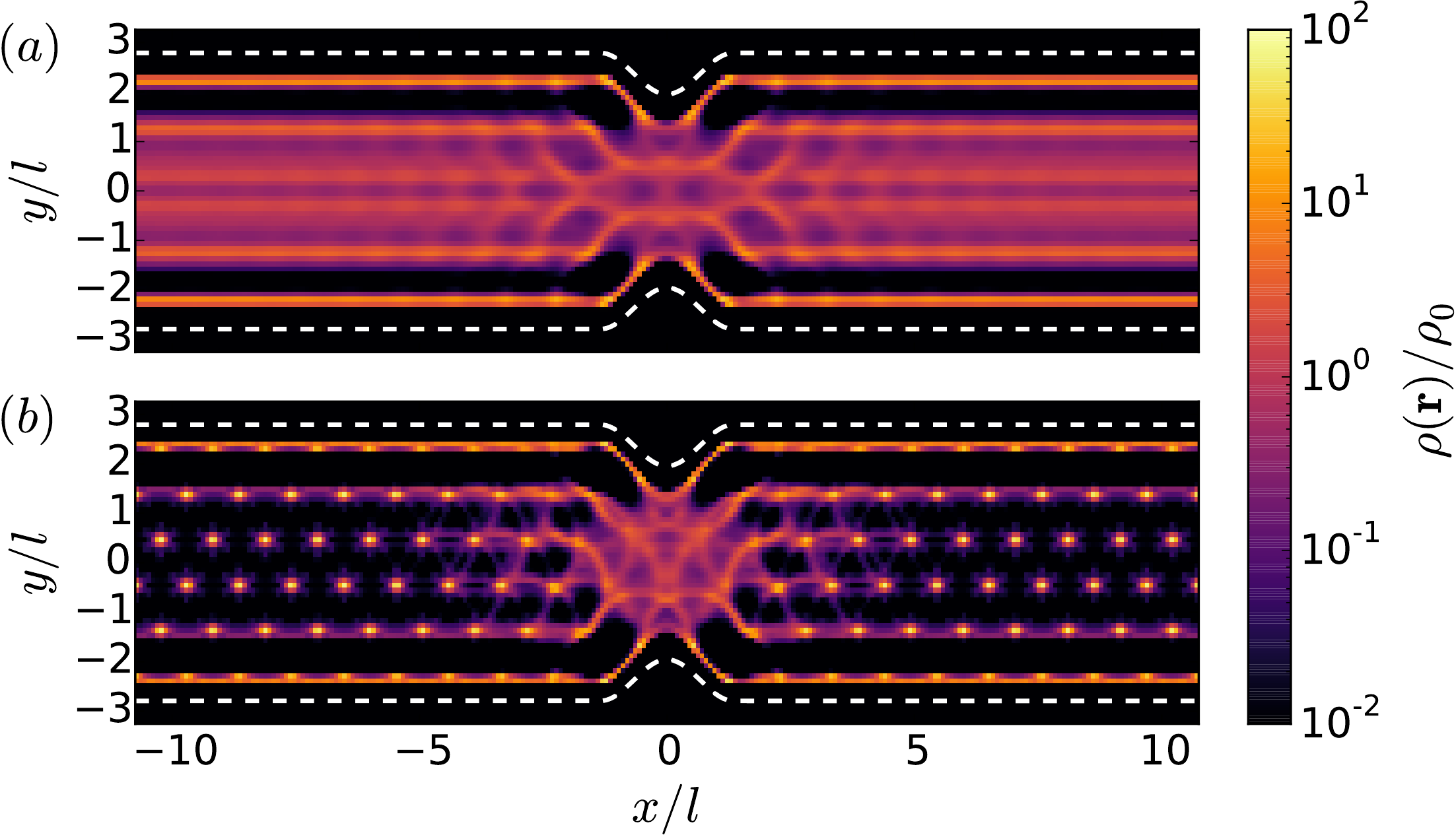}\\
\caption{Equilibrium density profiles $\rho^{(0)}(\mathbf{r})$ as obtained from DDFT calculations without dragging force ($f = 0$) for {\bf(a)} low interaction strength $\Gamma = 20$ (fluid) and {\bf(b)} high interaction strength $\Gamma = 60$ (solid) at $L_y = 6d$ and $b = 0.7$.}
\label{fig:equilibrium_profile}
\end{figure}

The overall length of the system is chosen as $L_x = 21.5 l$ with periodic boundary conditions along the $x$-direction. As a discretisation we used $N_x \times N_y = 256 \times 64$ gridpoints. With prescribed density $\rho_0$ we have about $N=113-120$ particles in our system, depending on the constriction width $b$.

Starting from several initial density profiles, we solve the DDFT equation without any driving force to obtain an equilibrium density profile $\rho^{(0)}(\mathbf{r})$. We confirmed that the equilibrium profile does not depend on the initial profile. Depending on the coupling parameter $\Gamma$ we either obtain an inhomogenous fluid (Fig. \ref{fig:equilibrium_profile}a) or a crystalline profile of hexagonal order (Fig. \ref{fig:equilibrium_profile}b). The one-dimensional crystal in channel confinement can be observed for $\Gamma \gtrsim 30$, for both investigated channel widths $L_y = 5d$ and $L_y = 6d$. 

For $t > 0$ we switch on the driving force $f$, initiating the flow through the constriction.
We solve Eq. \eqref{DDFT} numerically using a finite volume partial differential equation solver \cite{FiPy:2009}.

\section{Brownian Dynamics simulations}\label{sec:bd}
\begin{figure}
\vspace{0.5cm}
 \includegraphics[width=0.45\textwidth]{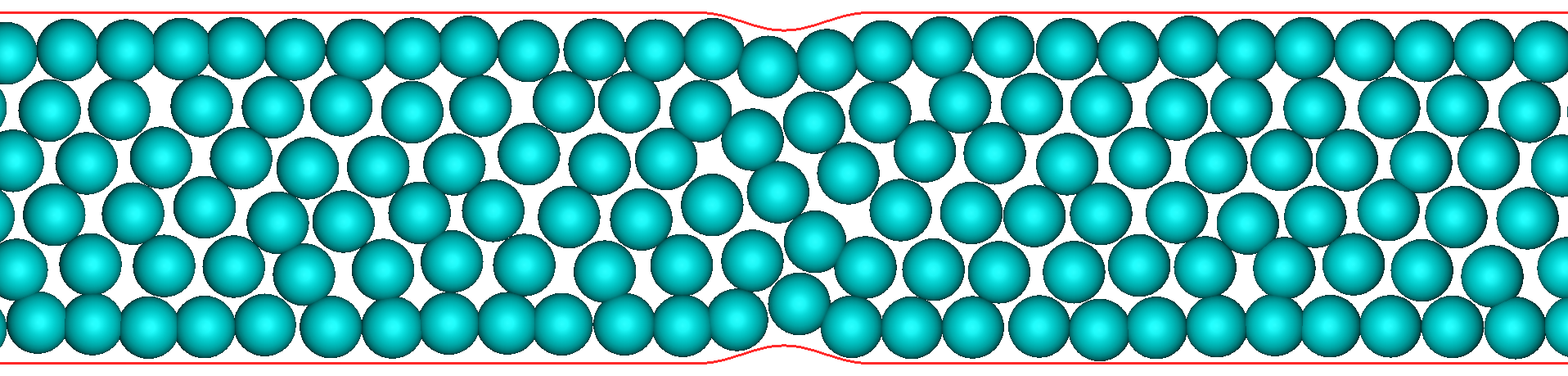}
\caption{Typical simulation snapshot of the system after equilibration without flow at $\Gamma = 20$, $L_y = 6d$, and $b = 0.9$.}
\label{fig:staticcrystal}
\end{figure}

In addition to DDFT calculations, we perform Brownian Dynamics simulations of the same system. In particular, we simulate $N=200$ particles with the same interparticle and particle-wall interactions as described above, using the equations of motion in Eq. \ref{eq:eom}. As in the DDFT calculations, we assume periodic boundary conditions along the $x$-direction.
In our simulations, we randomly place the particles into the channel, and let the system equilibrate in the absence of an external flow ($f = 0$). At sufficiently high interaction strength $\Gamma$, this typically results in a rapid ordering of the particles into a hexagonal crystal-like structure aligned with the confining walls. It should be noted that even in the absence of a constriction, this crystal is never defect-free: the two layers closest to the walls typically contain significantly more particles than those in the interior layers. This can be attributed to the long-ranged repulsion between the particles. Part of a typical snapshot of an equilibrated crystal is shown in Fig. \ref{fig:staticcrystal}. Larger defects (such as local square ordering) are occasionally observed at very high $\Gamma$, where the system can get trapped into a local energy minimum. However, these defects typically vanish rapidly once the flow is started.

Upon turning on the flow in the channel, the particles start moving (on average) in the direction of the flow. After an initial relaxation time, the flow through the channel reaches a steady state. In order to quantitatively examine the flow of particles in the channel, we directly measure the particle flux $j_x(x = x_0, t)$ through the constriction by counting in each timestep the number of particles passing through $x = x_0$. We average this flux over a large number ($\sim 10^4$) of runs. To do this, we run the simulation with flow for 100 $\tau$, then stop the flow and re-equilibrate the system first at a substantially lower effective interaction strength $\Gamma_\mathrm{relax} = \Gamma/10$ in order to allow for significant particle reorganization, and then re-equilibrate again at the original $\Gamma$. We then restart the flow and perform another measurement. Averaging over these runs, we obtain flow relaxation profiles for a range of combinations of $\Gamma$ and $b$.

\section{Results}\label{sec:results}

\begin{figure}
\includegraphics[width=0.4\textwidth]{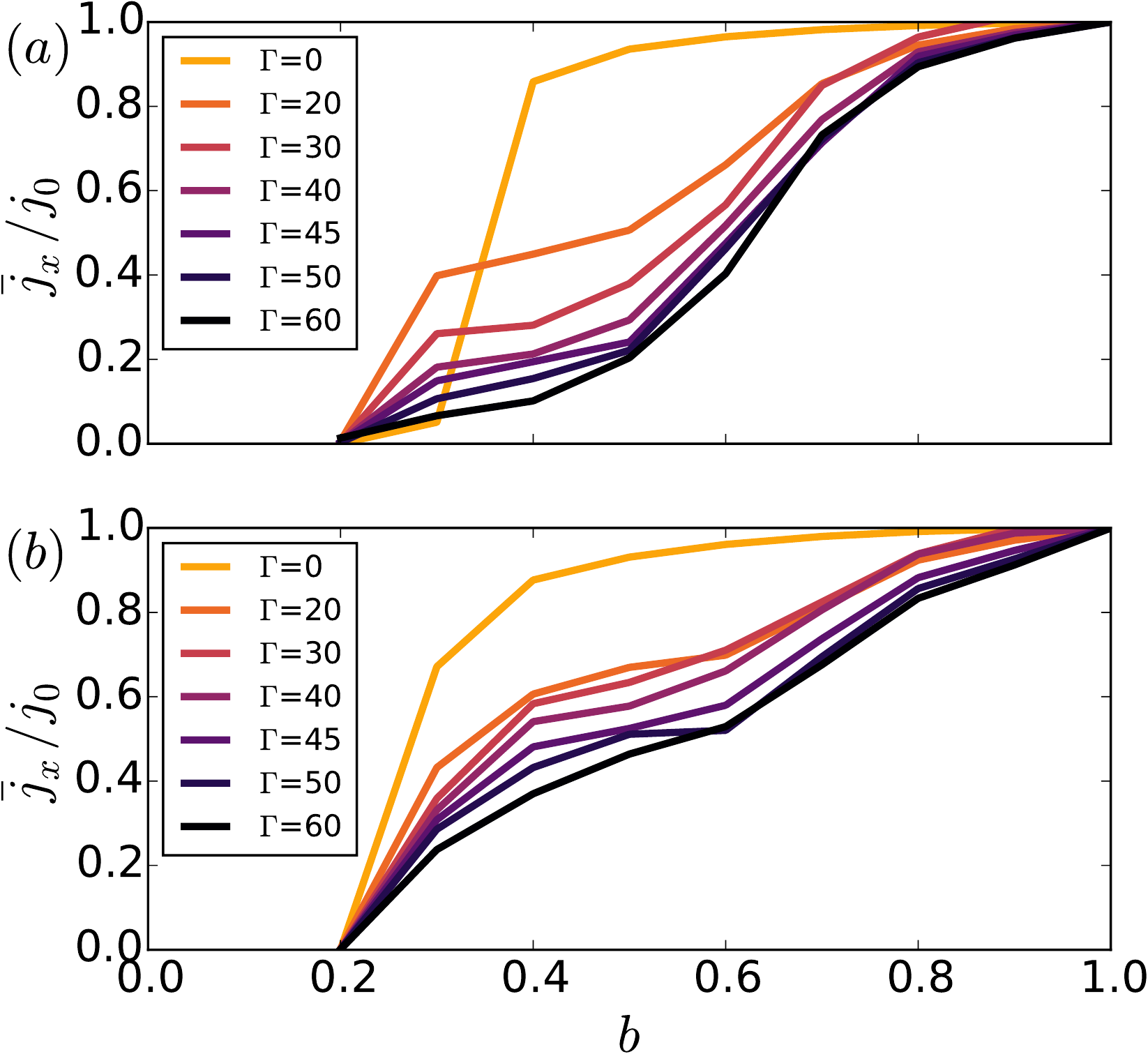}\\
\caption{Average particle flux $\bar{j}_x$ along the channel, as obtained from DDFT for channel widths {\bf (a)} $L_y = 5 d$, and {\bf (b)} $L_y = 6 d$. The flux is normalized by the average flux of an unconstricted system along the channel $j_0$. }
\label{fig:vplots_ddft}
\end{figure}

\begin{figure}
\includegraphics[width=0.4\textwidth]{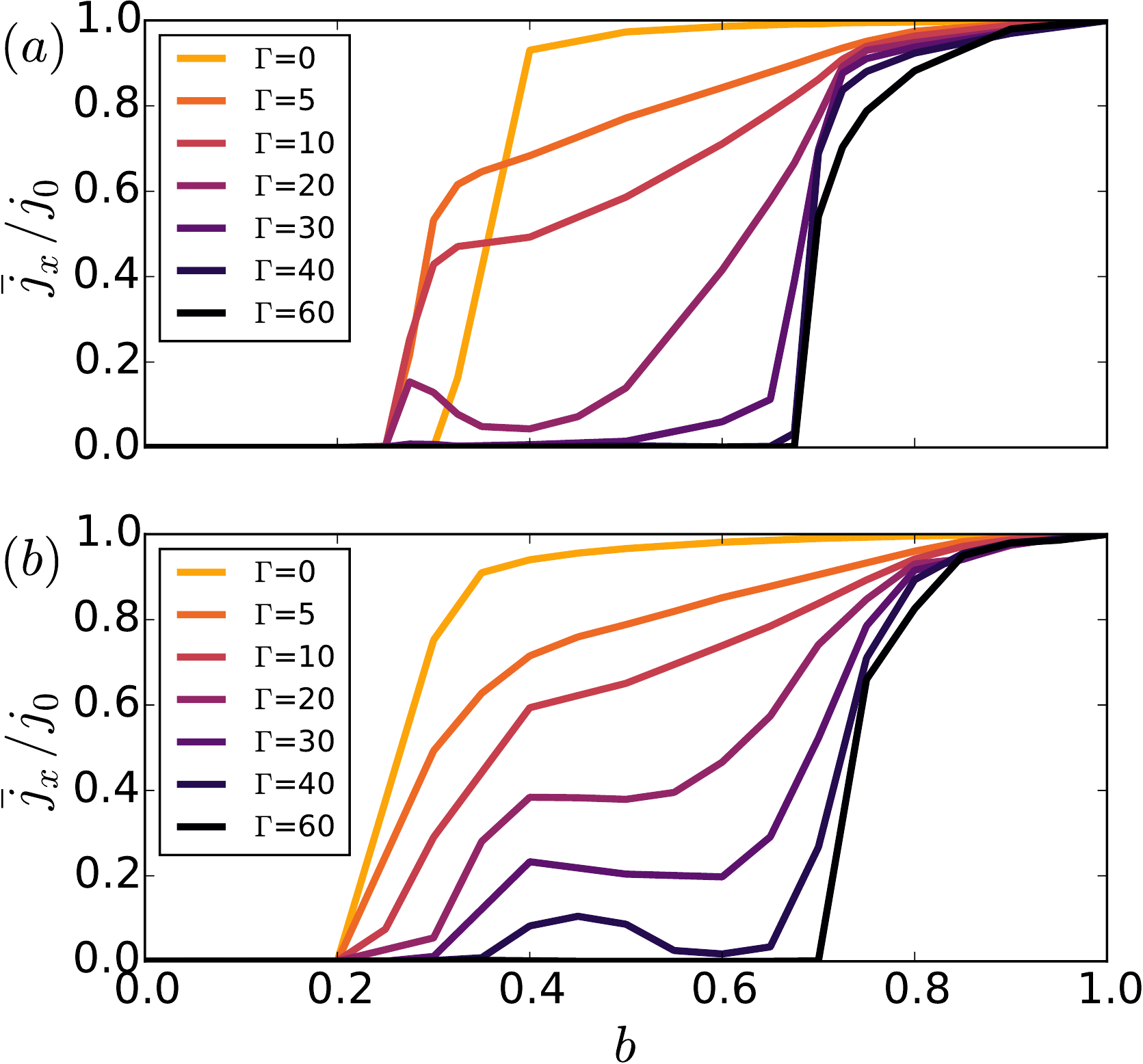}\\
\caption{Average particle flux $\bar{j}_x$ in the channel, as obtained from Brownian Dynamics simulations for channel widths {\bf (a)} $L_y = 5 d$, and {\bf (b)} $L_y = 6 d$. The flux is normalized by the average flux of free particles in an unconstricted channel $j_0$.    }
\label{fig:vplots}
\end{figure}

\subsection{Average flux}
The average flux in the system $\bar{j}_x$ for a range of coupling parameters $\Gamma$ and constriction widths $b$ is shown in figures \ref{fig:vplots_ddft} and \ref{fig:vplots}. In general, we observe that for stronger particle interactions the average flux is smaller, as the particles more effectively block each other from passing through the constriction. As expected, we also observe a decrease in average flux with decreasing constriction width $b$. We note, however, that in the simulations this trend is not always monotonic: there are regions where $\bar{j}_x$ decreases with increasing $b$. 

We observe qualitative agreement between the DDFT and simulation results. The main difference occurs at high $\Gamma$, where the simulations observe complete blocking ($\bar{j}_x = 0$), while the DDFT calculations predict a finite flux.
Additionally, the DDFT calculations predict only a monotonous increase in $\bar{j}_x$ with $b$ for the investigated parameter range.

It should be noted here that the observed results are expected to be influenced strongly by the length of the channel: at constant number density, a longer channel implies that the drag force $f$ is applied to a larger number of particles in front of the constriction. This results in a proportional increase in the pressure near the constriction, which is expected to enhance the flow of particles. Indeed, simulations on larger systems ($N=400$) and on systems with larger external forces confirm that doubling the channel length is approximately equivalent to doubling the external drag force on the particles.

\begin{figure}
\includegraphics[width=0.4\textwidth]{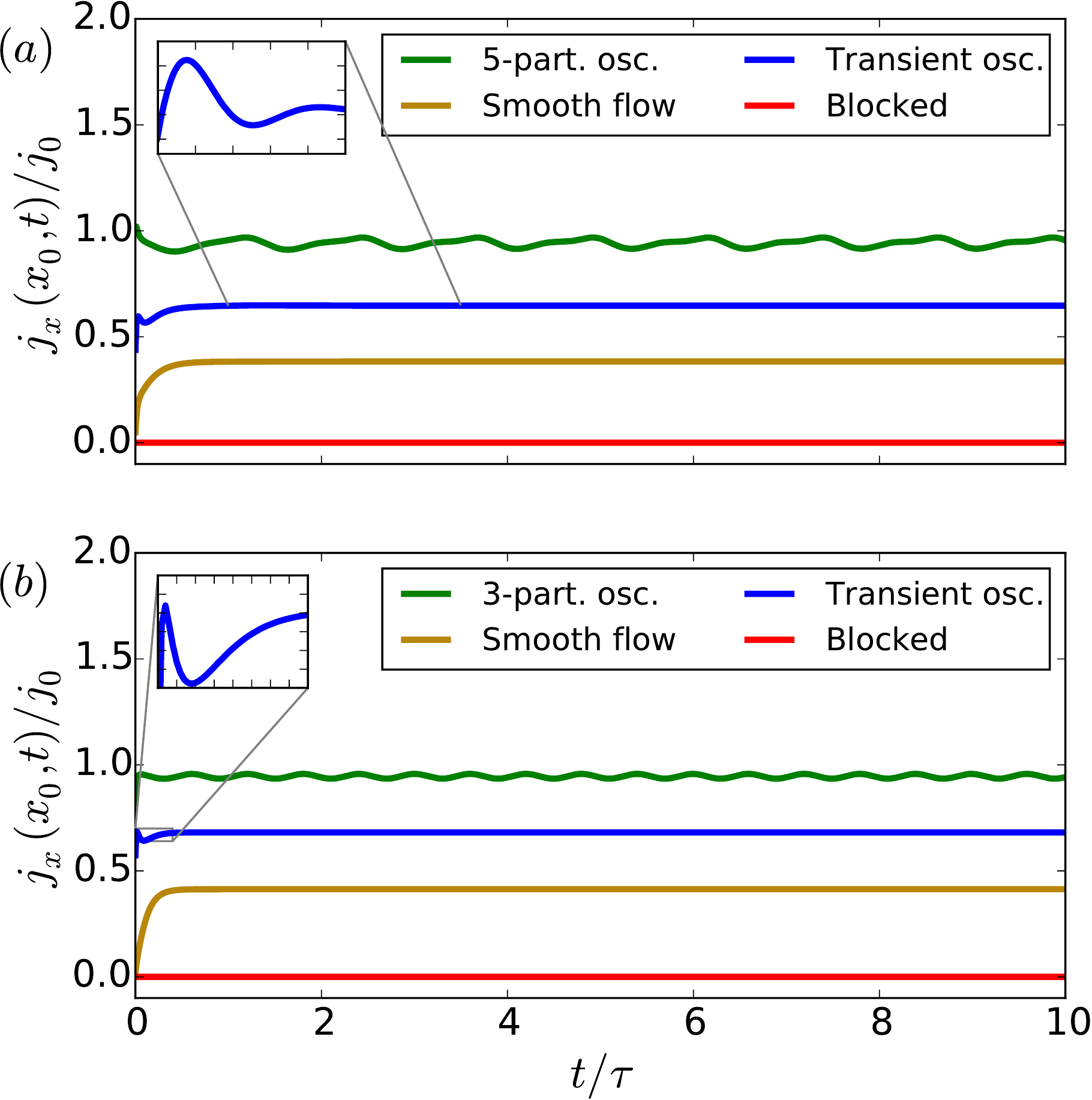}\\
\caption{Average particle flux $j_x(x_0,t)$ along the channel through the constriction as a function of time $t$ elapsed since starting the flow, as obtained from DDFT calculations for channel widths {\bf (a)} $L_y = 5d$, and {\bf (b)} $L_y = 6 d$ for $\Gamma=30, b=0.9$ (top, green) and $\Gamma=20$ with constriction width $b=0.6, 0.3$ and $0.2$ (bottom, red). These selected examples illustrate the different states as shown in \ref{fig:statediagrams}. The flux is normalized by its average value $j_0$ in an unconstricted channel (i.e. $\bar{j}_x$ at $b = 1$) at the same $f$. The inset shows a zoom of the particle flux in the transient state and highlights a weak and decaying oscillation.}
\label{fig:fluxplots_ddft}
\end{figure}

\begin{figure}
\includegraphics[width=0.4\textwidth]{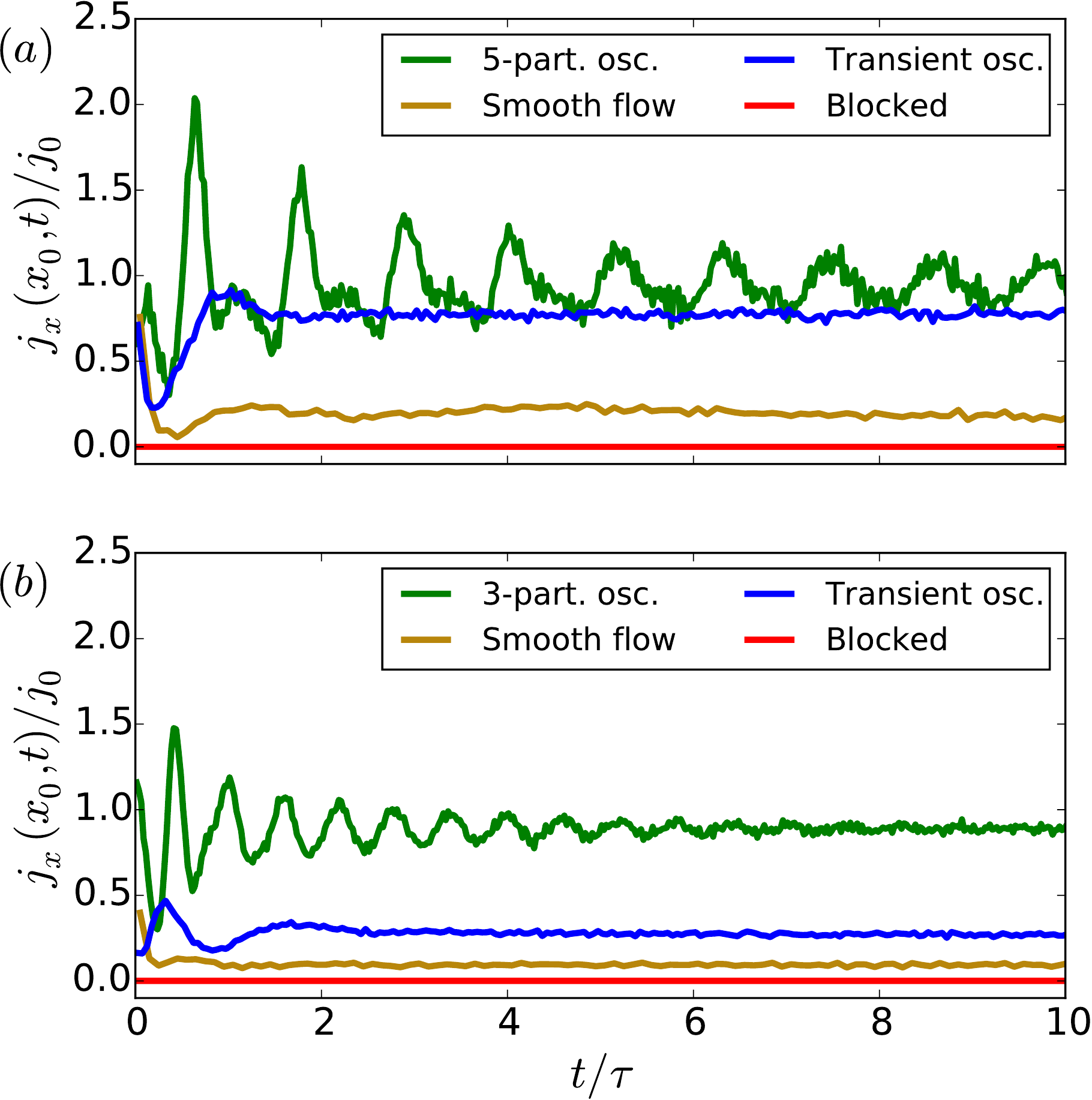}\\
\caption{Plots of the average particle flux $j_x(x_0,t)$ through the middle of the constriction as a function of time $t$ elapsed since starting the flow, as obtained from Brownian Dynamics simulations for channel widths {\bf (a)} $L_y = 5 d$, with $\Gamma = 20$, and {\bf (b)} $L_y = 6 d$, with $\Gamma = 40$. For both widths, the constriction widths are given by $b = 0.8$ (top, green), $0.7, 0.5$ and $0.2$ (bottom, red). The flux is normalized by its average value in an unconstricted channel (i.e. $\bar{j}_x$ at $b = 1$) at the same $f$.  }
\label{fig:fluxplots}
\end{figure}

\subsection{Flow behavior}
\begin{figure*}
\begin{tabular}{ll}
a) Simulations, 5 layers&b) Simulations, 6 layers\\
 \includegraphics[width=0.35\textwidth]{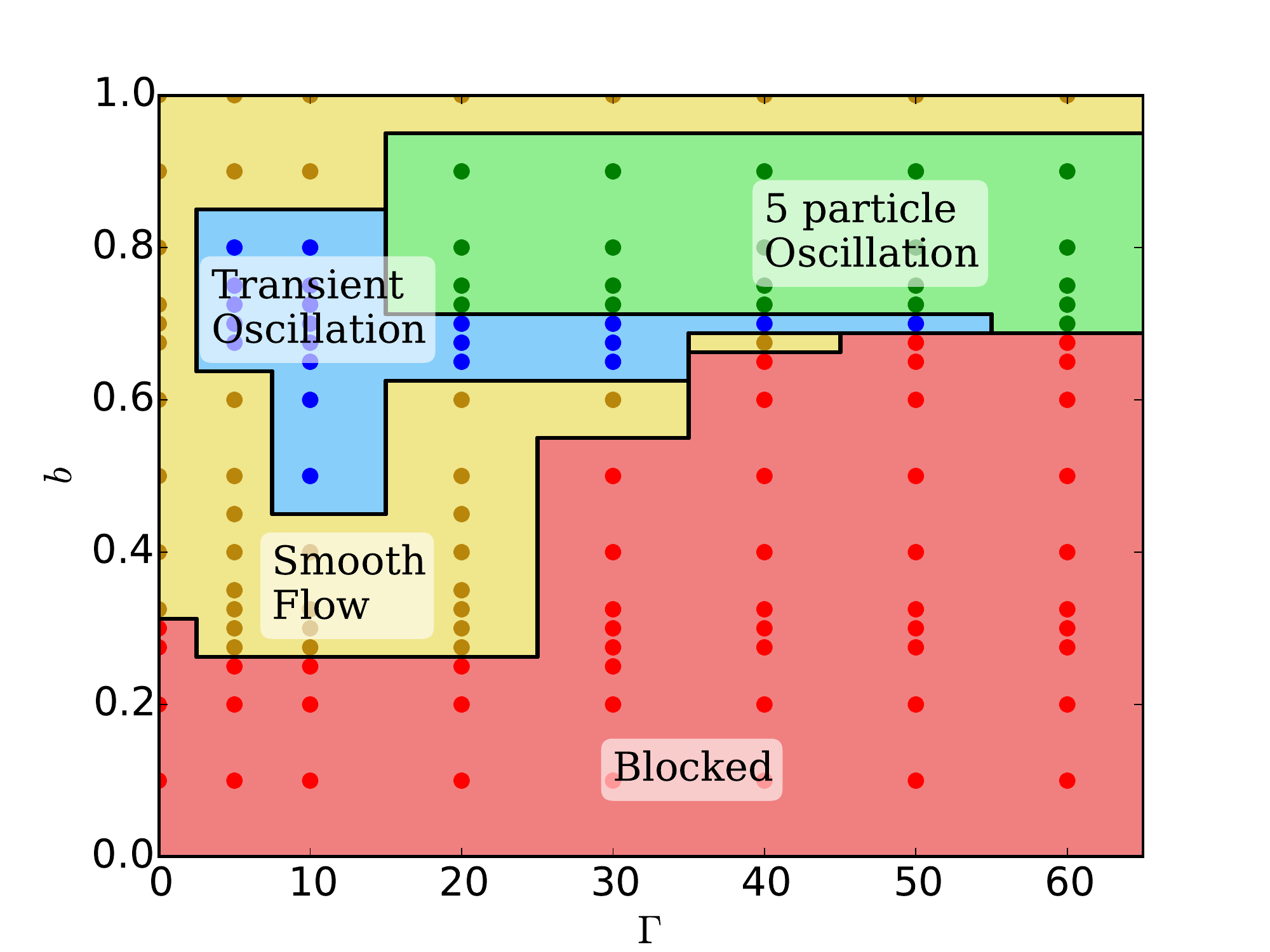}\hspace{1cm} & \includegraphics[width=0.35\textwidth]{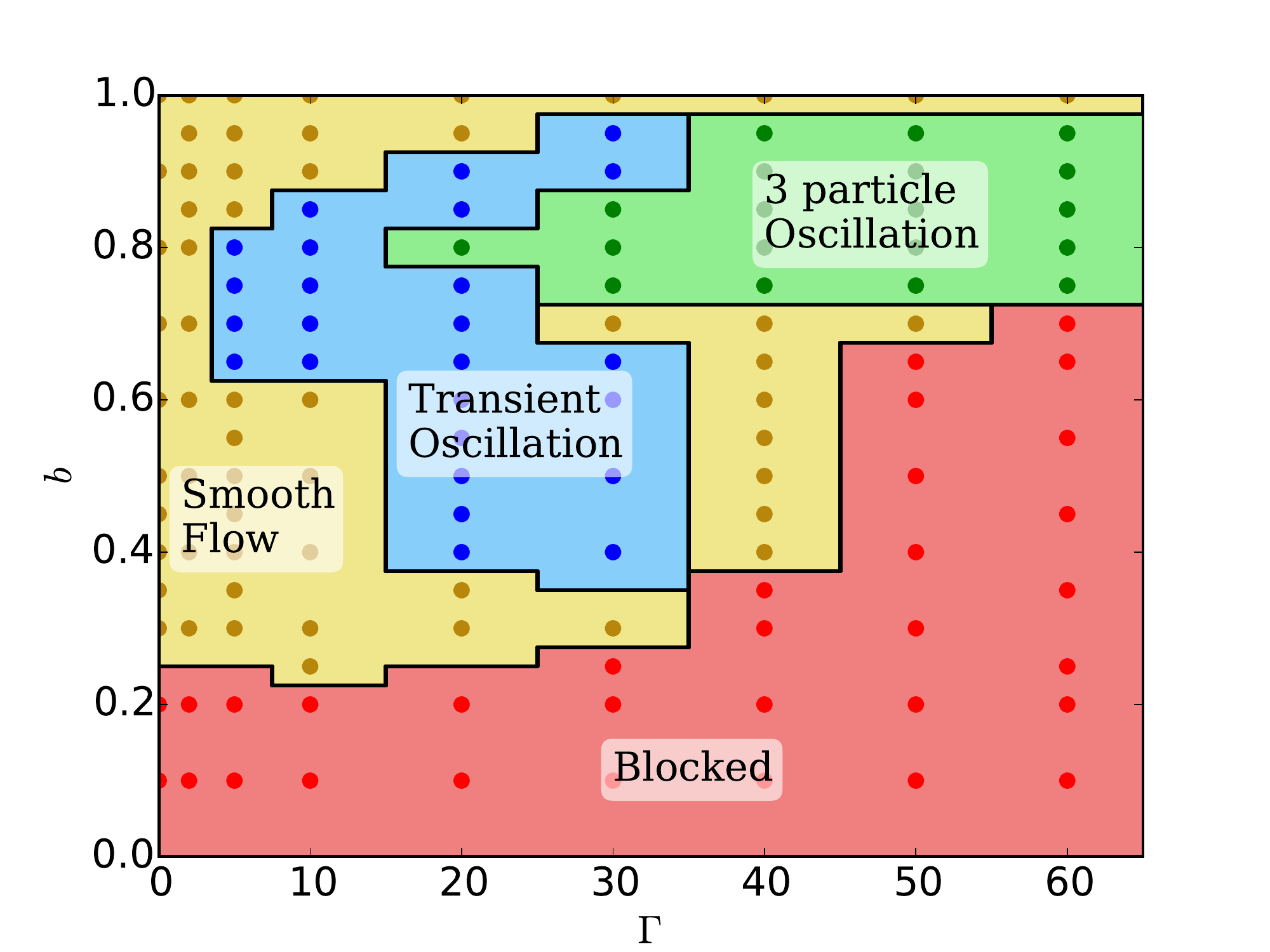} \\
c) DDFT, 5 layers&d) DDFT, 6 layers\\
 \includegraphics[width=0.35\textwidth]{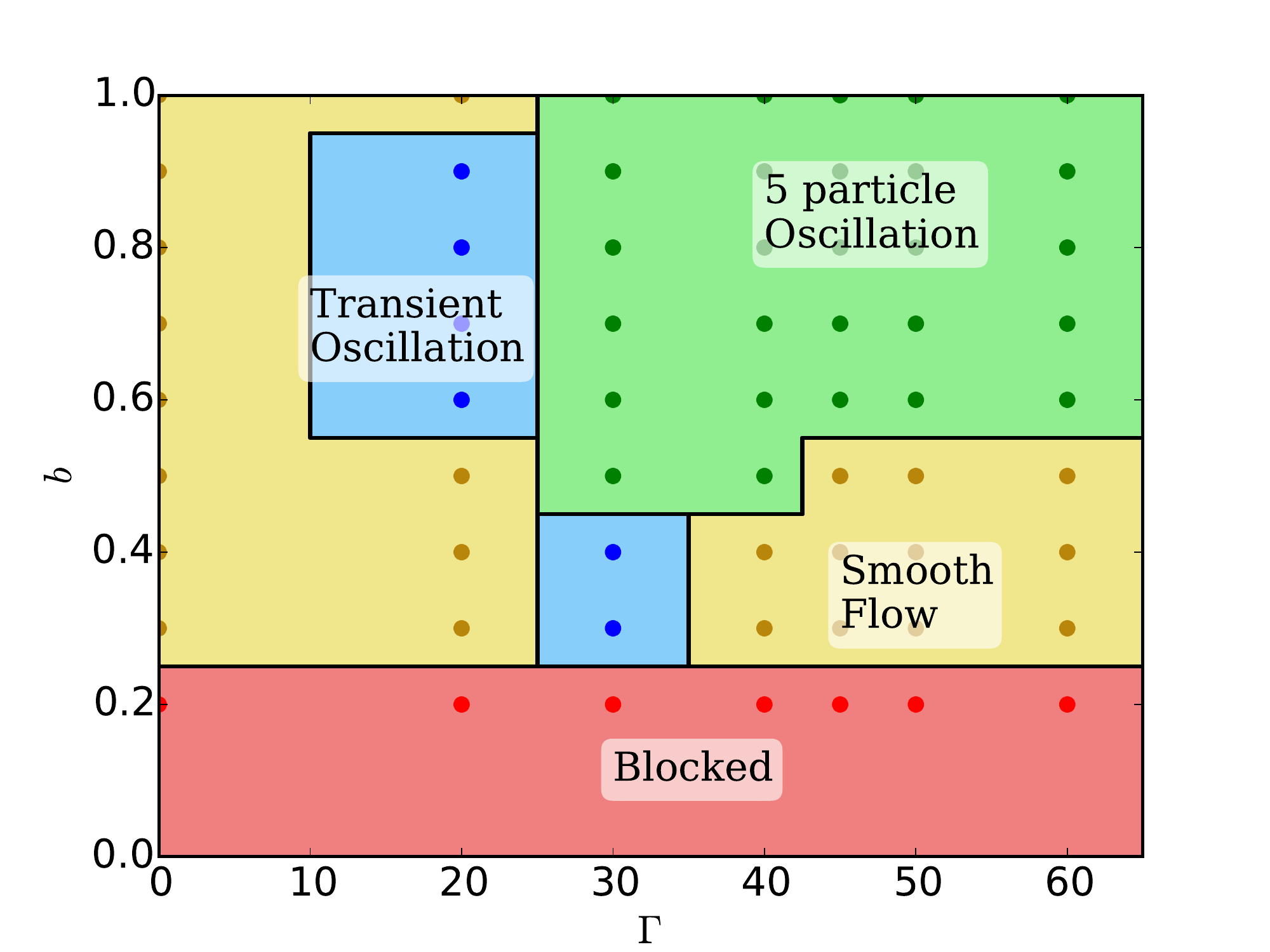}\hspace{1cm} & \includegraphics[width=0.35\textwidth]{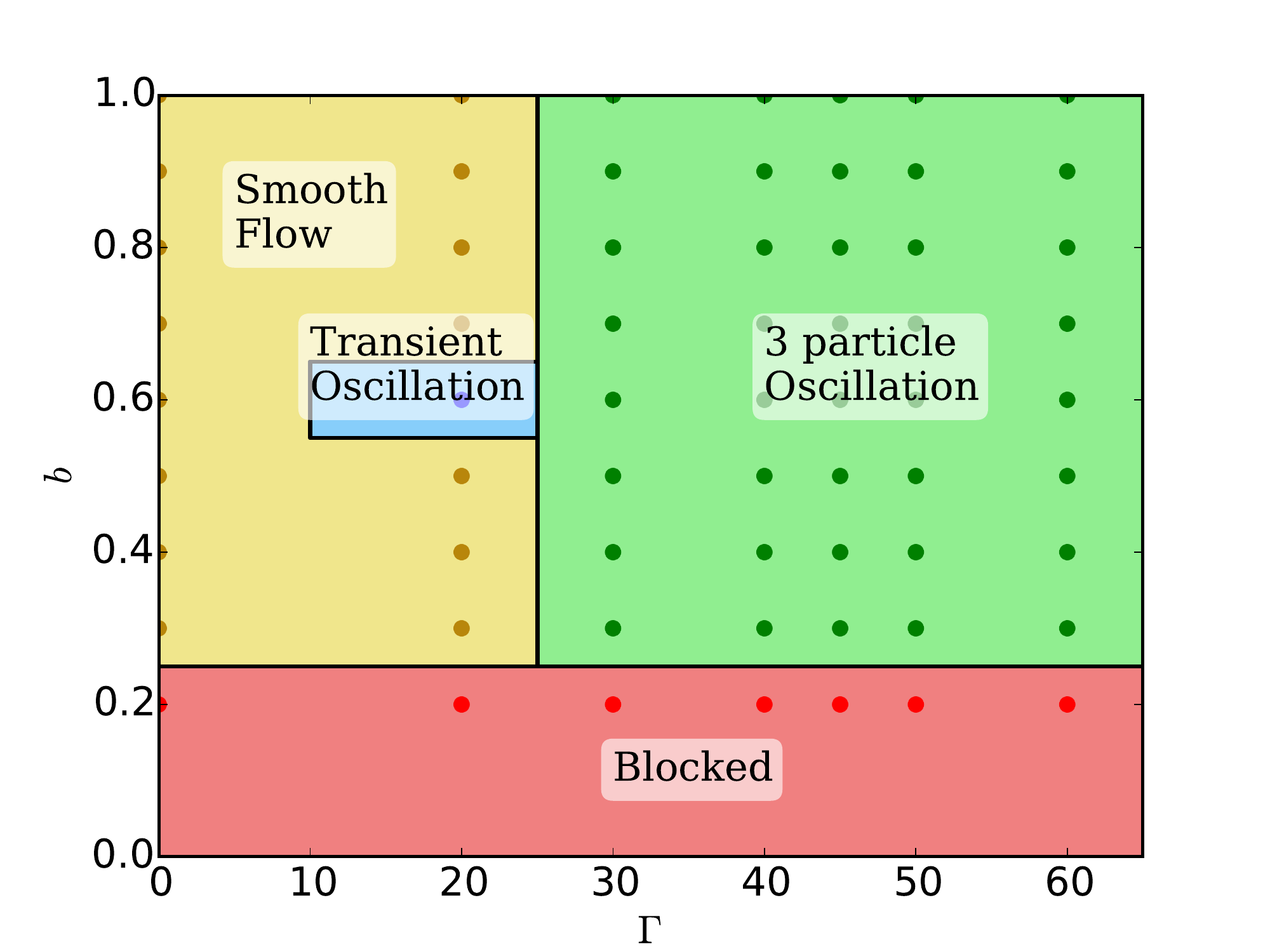} 
\end{tabular}
\caption{State diagrams indicating the types of flow observed for channels of width {\bf (a)}, {\bf (c)} $L_y = 5d$ and {\bf (b)} , {\bf (d)} $L_y = 6d$, as obtained from Brownian Dynamics simulations {\bf (a)}, {\bf (b)} and DDFT calculations {\bf (c)}, {\bf (d)}. In all cases $f = 1 k_B T/l$, and $N=200$ or $\bar{\rho} = \rho_0$, respectively. The dark colored points indicate points where DDFT calculation and simulations were performed.}
\label{fig:statediagrams}
\end{figure*}

After starting the flow, we observe four qualitatively different types of flow behavior in both our DDFT results and our simulations. First, we distinguish between systems that show a complete blockade (i.e. zero particle flow $\bar{j}_x$), and particles that show a finite flow of particles. In the case of a finite flow, the average flux through the constriction eventually reaches a constant value in the simulations. However, shortly after starting the flow, we often observe oscillations in the flux that decay over time. In this regime, we observe three types of decay: an almost immediate decay to a smooth flow, a brief period of transient oscillations without a clearly defined periodicity, and a long-time oscillation with a period which is independent of $b$ and $\Gamma$. In the DDFT calculations we observe the same regimes. However, due to the lack of stochastic noise, in the long-time oscillation regime, the DDFT calculations predict periodic (i.e. non-decaying) oscillations. Below, we discuss each type of flow in detail. In Figs. \ref{fig:fluxplots_ddft} and \ref{fig:fluxplots}, we plot the average flux through the constriction as a function of time as obtained from DDFT and simulations, respectively, for each of the four types of flow, and for channel widths $L_y = 5d$ and $L_y = 6d$. Additionally, in Fig. \ref{fig:statediagrams}, we show state diagrams for the same two channel widths from both simulations and DDFT, where we show the type of flow observed for a range of investigated values of $b$ and $\Gamma$.

\subsubsection{Blockade}
A blockade of the particle flow in the system (as observed on the time scale of the calculation)
occurs at narrow constrictions $b \lesssim 0.2$ for all $\Gamma$.  This is an effect of the softness of the confining 
potential $V_\text{ext}$. Due to this softness, a potential barrier on the order $k_B T$ starts 
appearing in the center of the channel around $b \simeq 0.3$ for both channel widths considered, which increases rapidly for smaller $b$.

For high particle interaction strengths $\Gamma \gtrsim 40$ an additional blockade situation for wider constrictions can be observed in the computer simulations. At sufficiently high $\Gamma$, the highly ordered lattice resists the deformations necessary to allow the flow of particles through the constriction. In the DDFT calculations, this effect is not observed probably due to the insufficient treatment of particle correlations within
the Ramakrishnan-Yussouff approximation.

\subsubsection{Smooth Flow}\label{sec:smoothflow}
This flow behavior is characterised by an overdamped transient flow that converges to a constant level almost immediately. It can be observed in the fluid phase at intermediate constriction widths. For larger $\Gamma$ values we can find the smooth flow behavior also in the 5 layer DDFT system and the 6 layer simulation system.

\subsubsection{Three or five-particle Oscillation}\label{sec:oscillation}
\begin{figure*}
\begin{tabular}{cc}
a)&\\
&\includegraphics[height=2.5cm]{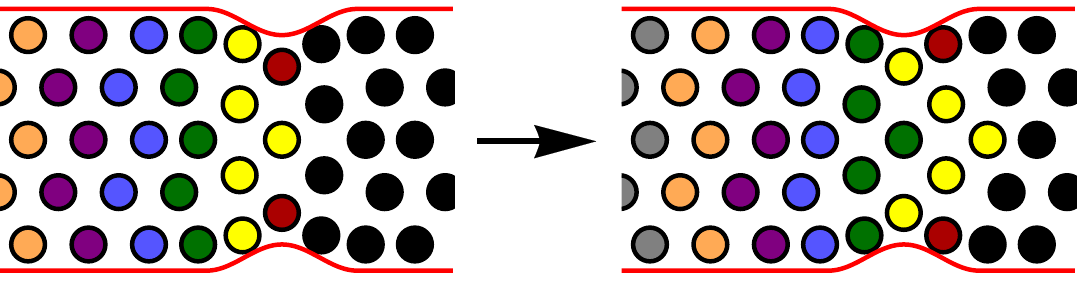}\\
b)&\\
&\includegraphics[height=3cm]{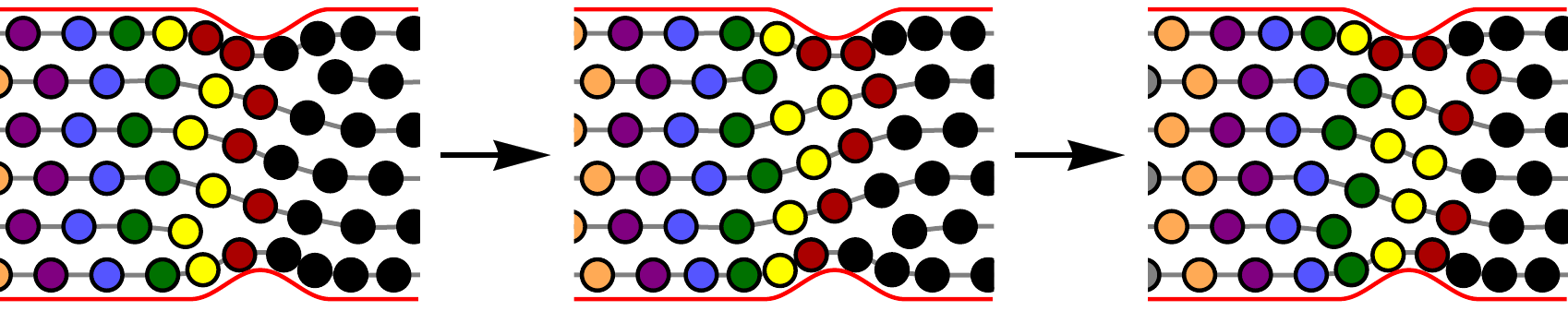}\\
\end{tabular}
\caption{Schematic picture of the periodic flow of the crystal observed for high interaction strength $\Gamma$ and wide constrictions (large $b$). The figures are idealized snapshots of the system separated in time by exactly one oscillation period. Flow is from left to right. {\bf (a)} For a channel width $L_y = 5 d$, five crystal layers form, and one oscillation corresponds to the movement of the crystal by one lattice spacing. During this time, each particle assumes the position of the particle in front of it. {\bf (b)} For a crystal with six layers ($L_y = 6d$), one oscillation period corresponds to the movement of the crystal by half of a lattice spacing. Note that in the case of six layers, the up-down symmetry in the system is broken, and we observe two symmetric dislocations in the crystal pattern, as indicated by the gray lattice lines. After one oscillation, the locations of the particles (and dislocations) are vertically mirrored with respect to the initial configuration (middle snapshot). After the next oscillation (right snapshot), we recover the original configuration. Note that for both channel widths, the higher concentration of particles in the top and bottom layers of the crystal results in a lower velocity of the particles in those layers.}
\label{fig:crystalcartoons}
\end{figure*}

For intermediate to strong particle interactions and for intermediate to wide constrictions we observe strong oscillatory behavior in the 
particle flow. While in the Brownian Dynamics simulations the oscillation is damped we can find for the DDFT results an undamped oscillation that is periodic after a brief transient phase. This can be 
understood from the fact that the damping in the simulation is due to the presence of fluctuations, which are missed in the mean-field 
approach of the DDFT. We expect that the fluctuations which destroy long-ranged periodic order in one dimensions are also
responsible for washing out the correlations in the flow dynamics. The frequency of the oscillation depends on the number of particle layers in the system. For a five layer system the 
frequency is lower and corresponds to five particles passing the constriction during one oscillation period. In contrast, in the six layer 
system we observe a higher frequency in the flow oscillation, corresponding to three particles passing the constriction. In Fig. \ref{fig:crystalcartoons} we illustrate the mechanism that is responsible for this qualitative difference. For both channel widths, the oscillation period represents the smallest number of particles that can pass through the constriction in such a way that the system reverts to its original configuration. In the case of an odd number of layers (i.e. five), this is simply five particles, such that the crystalline lattice shifts by one lattice spacing. For an even number of crystalline layers (i.e. $L_y = 6d$), this period instead corresponds to three particles passing through the constriction, such that the lattice moves by half a lattice spacing, and then coincides with a vertically mirrored version of the initial lattice. We have confirmed with simulations that the same mechanism occurs for other (small) numbers of layers.

In the supplementary material we include two movies of these dynamics in a system with $L_y = 6d$ and $b=0.8$ as obtained from DDFT and simulations.

As can be seen in Fig. \ref{fig:statediagrams}, in our DDFT findings this type of oscillatory flow is dominant in a significantly larger region of the ($b$, $\Gamma$) parameter space. In the simulations, this mechanism only occurs around $b \simeq 0.8$. Likely, this can be attributed to the approximative excess functional which cannot account for complex crystal configurations that is responsible for the blockade in front of the constriction.

\subsubsection{Transient Oscillation}\label{sec:transient}
In addition to the overdamped decay to a smooth flow and the long-time mechanism described above, we also observe short transient fluctuations that converge to a constant level within a few oscillations. Unlike the smooth flow the transient regime is not overdamped but performs several oscillations around the final level. It can be found for intermediate particle interaction strengths and wide constrictions. Note that while these transient oscillations are clearly distinguishable from the long-term fluctuations described above via their period, the distinction between the overdamped decay to a smooth flow and these transient fluctuations are often less clear. In particular, in the simulations, the presence of statistical noise makes determining the presence of secondary or tertiary peaks in the flow profile difficult if their amplitude is small. Due to the absence of statistical noise in the DDFT calculations transient fluctuations are better distinguishable from the smooth flow.

\section{Conclusions}\label{sec:conclusion}
In conclusion, we have explored the flow of two-dimensional solids and fluids through geometric 
constrictions on a particle-resolved level by using
models describing the Brownian dynamics of strongly interacting colloids
in a linear channel. Upon starting the flow, four different situations were identified
using dynamical density functional theory
and particle-resolved computer simulations: i) a complete blockade,
ii) a smooth flow, iii) an oscillatory behaviour in the particle flux,
iv) a long-lived stop-and-go behaviour in the flow.
Though the dynamical density functional theory is an
approximative mean-field theory, it qualitatively describes the most of the states and trends.

Our predictions can be confirmed by using magnetic colloidal particles driven through microchannels 
\cite{KreuterSNLE2013,Siems_SR_2012}
as already used for the flow over energetic barriers but in the absence of constrictions \cite{Kreuter_JPCM_2012}.
For this realization,
flow and diffusion through linear channels involving $4-8$ layers
 has been considered before \cite{Koeppl_PRL_2006,Henseler_PRE_2010,Wilms_PRE_2012}
and a layer reduction was found.
However, an  extreme
geometric narrowing in the channel was not studied in  previous work.
But it can be done by using micropatterned channels \cite{GenoveseS2011}.

Future work should address three-dimensional constrictions
(like an colloidal hour-glass) although clearly the numerical
evaluation of DDFT in three dimensions is harder. It would be
nice to explore colloidal mictures driven through constrictions \cite{GlanzWittkowskiL_2015}
where we expect a rich scenario of flow states depending
on the microscopic interactions.

We note that in our model the constriction
 was seen only by the colloids only but not by the solvent.
Such barriers can be prepared using laser-optical forces
which only act on the colloids but are invisible by the solvent,
 i.e. they allow for a
full solvent penetration. Real geometric constrictions
governed by the shape of the channel also affect the solvent flow. 
The same is true when the flow is generated by a pressure gradient in the solvent
These situations require a more detailed modelling regarding
 the solvent flow field
which provides additional advective drag forces to the colloids.
For a single particle moving through a constriction, the solvent effect was taken into accout by Martens and coworkers
\cite{Martens_PRL_2013,Martens_EPJST_2014},
 for another situation see Ref. \cite{Malgaretti_PRL_2014}.
 More realistic calculations which include the hydrodynamics of the solvent and
 the hydrodynamic interactions between the colloids are still be be done in future studies.

\section*{Acknowledgements}
This work was financially supported by the ERC Advanced Grant INTERCOCOS (Grant No. 267499). FS acknowledges support from
the Alexander-von-Humboldt foundation.

\bibliography{constriction_references}

\end{document}